\documentclass[letterpaper, 10 pt, conference]{ieeeconf}  % Comment this line out if you need a4paper

\IEEEoverridecommandlockouts                              % This command is only needed if 
\usepackage{graphics} % for pdf, bitmapped graphics files
\usepackage{epsfig} % for postscript graphics files
\usepackage{amsmath} % assumes amsmath package installed
\usepackage{cleveref}
\usepackage{tikz}\usetikzlibrary{circuits.ee.IEC}
\usepackage{algorithm}
\usepackage{algorithmic}
\usepackage{cite}

\title{\LARGE \bf
Frequency Response of Transmission Lines with Unevenly Distributed Properties with Application to Railway Safety Monitoring*
}

\author{Xiangyu Ni$^{1}$ and Bill Goodwine$^{2}$% <-this % stops a space
\thanks{*The support of the US National Science Foundation under Grant No. CMMI 1826079 is gratefully acknowledged.}% <-this % stops a space
\thanks{$^{1}$Xiangyu Ni is with {\tt\small xni@nd.edu}}%
\thanks{$^{2}$Bill Goodwine is with {\tt\small billgoodwine@nd.edu}}%
}

\begin{document}

\maketitle

%%%%%%%%%%%%%%%%%%%%%%%%%%%%%%%%%%%%%%%%%%%%%%%%%%%%%%%%%%%%%%%%%%%%%%%%%%%%%%%%
\begin{abstract}

This paper proposes a method to quickly and efficiently compute the voltage and current along a transmission line which can be ''damaged''; that is its electrical properties can be unevenly distributed. The method approximates a transmission line by a self-similar circuit network and leverages our previous work regarding the frequency response for that class of networks. The main motivation arises from the research for railway track circuit systems where transmission line models are often employed. However, in contrast to real transmission lines, railway track circuits are more likely to be damaged due to its scale and environmental uncertainties; furthermore, changes in circuit properties due to a train occupying a segment of the track also is of great interest as a means to ensure safety. As a result, an accurate and quick simulation of damaged track circuits is necessary and can contribute to the corresponding health monitoring research area in the future.

\end{abstract}

%%%%%%%%%%%%%%%%%%%%%%%%%%%%%%%%%%%%%%%%%%%%%%%%%%%%%%%%%%%%%%%%%%%%%%%%%%%%%%%%
\section{INTRODUCTION}
Transmission line theory, credited to Oliver Heaviside, determines the voltage and current along a transmission line with respect to both spatial and temporal parameters. Its effect is significant for the wiring purpose especially when dealing with high frequency signals \cite{Wilson2012Grounding}. It is widely applied in electrical engineering. For antenna systems, it is used to determine the matching networks balancing a load with its source \cite{Rouphael2014Antenna}. It builds the foundation for physical models of power line communications channel representing signal propagation effects \cite{Lampe2016Power}. It is also employed to analyze the cylindrical body model for studying interaction of a human with electromagnetic fields \cite{Poljak2019Simplified}.

The main motivation of this work is the implementation of transmission line theory in railway track circuits which automatically detects whether a sector of track is occupied \cite{Hill1993Rail,Wang2016Fault,Zhao2009The,Verbert2015Exploiting}. One drawback of applying the classical transmission line theory to this problem is its assumption of uniformly distributed electrical properties, which is unlikely for track circuits. For instance, humidity in both soil and ballast bed can impact those properties as indicated by \cite{Hill1989Railway}. In addition, some sudden external influences, like lightning, can cause damages to a track circuit system which may lead to a catastrophic safety monitoring failures where two trains are present within the same track segment. 

As a result, in this work, we propose a method to quickly compute the voltage and current along a transmission line when its electrical properties are unevenly distributed with example applications to simulating a damaged track circuit system, as well as simulating a train passing along an intact track circuit. The contribution of this paper can be further employed more generally in the health monitoring research area for track circuits.

The proposed method approximates a transmission line by an electrical network with many subsections, and the electrical properties are lumped in each subsection. The electrical properties at one subsection can be different from the others, which imitates a transmission line with unevenly distributed properties. The transmission line model and its approximated counterpart, the circuit network model, are shown in \Cref{fig:transModel,fig:fullModel}, and are similar to the networks constructed in \cite{Zhao2009The,Chen2008Fault,Wang2010Modeling}. 

Leveraging a frequency-domain network modeling algorithm proposed in our previous work \cite{Ni2020Frequency}, we can obtain the impedance $V_g/I_g$ and the voltage gain $V_{out}/V_g$ at each subsection $g$. Then, given one voltage observation inside the network, \textit{e.g} $v_{out}$, we can obtain the voltage and current at every node connecting two adjacent subsections. Note that the network modeling algorithm presented in our previous work \cite{Ni2020Frequency} can be applied to any self-similar one-dimensional networks. Therefore, the network model is not limited to \Cref{fig:fullModel}, which is selected here merely due to its consistency with the transmission line model in \Cref{fig:transModel}. Readers with different circuit network models can still follow the same procedure proposed in this paper as well as the modeling algorithm in \cite{Ni2020Frequency}. This modeling approach yields a frequency domain model where damage (or the fact that a rail segment is occupied) is represented as a multiplicative disturbance, which is particularly convenient for robust control analyses. 

\begin{figure*}
    \centering
    \begin{tikzpicture}[circuit ee IEC]
        \draw (0,0) circle (2pt);
        \draw (0,-1.5) circle (2pt);
        \node at (-0.4,0) {$+$};
        \node at (-0.4,-1.5) {$-$};
        \node at (0,-0.75) (v1) {$v(x+\Delta x,t)$};
        \draw[thin, -latex] (v1.north) -- (0,-0.1);
        \draw[thin, -latex] (v1.south) -- (0,-1.4);
        \draw (0.8,0) to [resistor] node [midway,above=1pt] {$\frac{R}{2}\Delta x$} (2,0);
        \draw (2,0) to [inductor] node [midway,above=1pt] {$\frac{L}{2}\Delta x$} (3.2,0);
        \draw[thin, -latex] (1.6,-0.3) -- node [midway,below=0.1] {$i(x+\Delta x,t)$} (2.4,-0.3);
        \draw (0.8,-1.5) to [resistor] node [midway,above=1pt] {$\frac{R}{2}\Delta x$} (2,-1.5);
        \draw (2,-1.5) to [inductor] node [midway,above=1pt] {$\frac{L}{2}\Delta x$} (3.2,-1.5);
        \draw[thin] (0.1,0) -- (0.8,0);
        \draw[thin] (0.1,-1.5) -- (0.8,-1.5);
        \draw (3.2,0) to [resistor] node [midway,right=3pt] {$G\Delta x$} (3.2,-1.5);
        \draw (4.5,0) to [capacitor] node [midway,right=5pt] {$C\Delta x$} (4.5,-1.5);
        \draw[thin, -latex] (5.7,-0.3) -- node [midway,below=0.1] {$i(x,t)$} (6.5,-0.3);
        \node at (7.2,-0.75) (v2) {$v(x,t)$};
        \draw (7.2,0) circle (2pt);
        \draw (7.2,-1.5) circle (2pt);
        \draw[thin] (3.2,0) -- (7.1,0);
        \draw[thin] (3.2,-1.5) -- (7.1,-1.5);
        \draw[thin, -latex] (v2.north) -- (7.2,-0.1);
        \draw[thin, -latex] (v2.south) -- (7.2,-1.4);
        \node at (3.6,-2) (dx) {$\Delta x$};
        \draw[thin, -latex] (dx.west) -- (0,-2);
        \draw[thin] (dx.east) -- (7.2,-2);
        \draw[thin] (0,-1.7) -- (0,-2.3);
        \draw[thin] (7.2,-1.7) -- (7.2,-2.3);
        \draw[thin] (7.3,0) -- (7.7,0);
        \draw[thin] (7.3,-1.5) -- (7.7,-1.5);
        \node at (8.1,0) {$\cdots$};
        \node at (8.1,-1.5) {$\cdots$};
        \draw[thin, -latex] (8.5,-0.3) -- node [midway,below=0.1] {$i(0,t)$} (9.3,-0.3);
        \draw (9.7,0) to [resistor] node [midway,right=3pt] {$Z_0$} (9.7,-1.5);
        \node at (11,-0.75) (v0) {$v(0,t)$};
        \draw[thin, -latex] (v0.north) -- (11,-0.1);
        \draw[thin, -latex] (v0.south) -- (11,-1.4);
        \draw (11,0) circle (2pt);
        \draw (11,-1.5) circle (2pt);
        \draw[thin] (8.5,0) -- (10.9,0);
        \draw[thin] (8.5,-1.5) -- (10.9,-1.5);
        \draw[thin] (9.7,-1.7) -- (9.7,-2.3);
        \node at (8.45,-2) (x) {$x$};
        \draw[thin, -latex] (x.west) -- (7.2,-2);
        \draw[thin] (x.east) -- (9.7,-2);
    \end{tikzpicture}
    \caption{Model for transmission line theory. The left hand side is the input/transmitter end. The right hand side is the output/receiver end, where $x=0$}
    \label{fig:transModel}
\end{figure*}
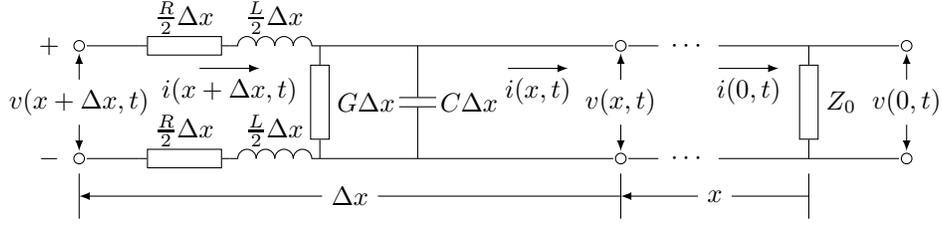
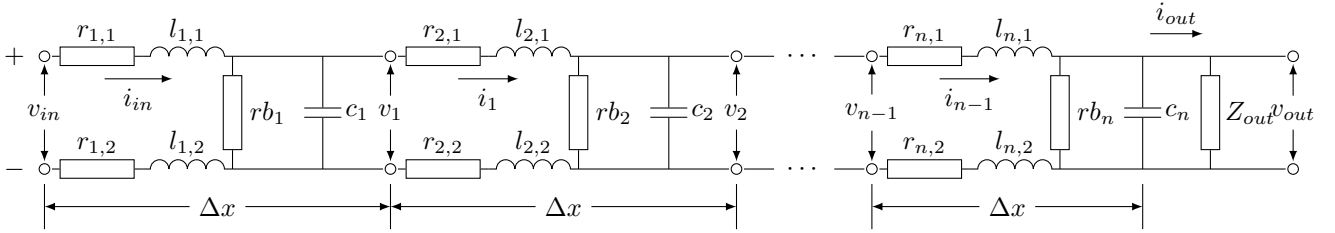
\begin{figure*}
    \centering
    \begin{tikzpicture}[circuit ee IEC]
        \draw (0,0) circle (2pt);
        \draw (0,-1.5) circle (2pt);
        \node at (-0.4,0) {$+$};
        \node at (-0.4,-1.5) {$-$};
        \node at (0,-0.75) (vin) {$v_{in}$};
        \draw[thin, -latex] (vin.north) -- (0,-0.1);
        \draw[thin, -latex] (vin.south) -- (0,-1.4);
        \draw (0.1,0) to [resistor] node [midway,above=1pt] {$r_{1,1}$} (1.3,0);
        \draw (1.3,0) to [inductor] node [midway,above=1pt] {$l_{1,1}$} (2.5,0);
        \draw (0.1,-1.5) to [resistor] node [midway,above=1pt] {$r_{1,2}$} (1.3,-1.5);
        \draw (1.3,-1.5) to [inductor] node [midway,above=1pt] {$l_{1,2}$} (2.5,-1.5);
        \draw[thin, -latex] (0.8,-0.3) -- node [midway,below=0.1] {$i_{in}$} (1.7,-0.3);
        \draw (2.5,0) to [resistor] node [midway,right=3pt] {$rb_1$} (2.5,-1.5);
        \draw (3.7,0) to [capacitor] node [midway,right=5pt] {$c_1$} (3.7,-1.5);
        \node at (4.6,-0.75) (v1) {$v_1$};
        \draw (4.6,0) circle (2pt);
        \draw (4.6,-1.5) circle (2pt);
        \draw[thin, -latex] (v1.north) -- (4.6,-0.1);
        \draw[thin, -latex] (v1.south) -- (4.6,-1.4);
        \draw[thin] (0,-1.7) -- (0,-2.3);
        \draw[thin] (4.6,-1.7) -- (4.6,-2.3);
        \node at (2.3,-2) (dx1) {$\Delta x$};
        \draw[thin, -latex] (dx1.west) -- (0,-2);
        \draw[thin, -latex] (dx1.east) -- (4.6,-2);
        \draw[thin] (2.5,0) -- (4.5,0);
        \draw[thin] (2.5,-1.5) -- (4.5,-1.5);
        \draw (4.7,0) to [resistor] node [midway,above=1pt] {$r_{2,1}$} (5.9,0);
        \draw (5.9,0) to [inductor] node [midway,above=1pt] {$l_{2,1}$} (7.1,0);
        \draw (4.7,-1.5) to [resistor] node [midway,above=1pt] {$r_{2,2}$} (5.9,-1.5);
        \draw (5.9,-1.5) to [inductor] node [midway,above=1pt] {$l_{2,2}$} (7.1,-1.5);
        \draw[thin, -latex] (5.5,-0.3) -- node [midway,below=0.1] {$i_{1}$} (6.3,-0.3);
        \draw (7.1,0) to [resistor] node [midway,right=3pt] {$rb_2$} (7.1,-1.5);
        \draw (8.3,0) to [capacitor] node [midway,right=5pt] {$c_2$} (8.3,-1.5);
        \node at (9.2,-0.75) (v2) {$v_2$};
        \draw[thin, -latex] (v2.north) -- (9.2,-0.1);
        \draw[thin, -latex] (v2.south) -- (9.2,-1.4);
        \draw[thin] (9.2,-1.7) -- (9.2,-2.3);
        \node at (6.9,-2) (dx2) {$\Delta x$};
        \draw[thin, -latex] (dx2.west) -- (4.6,-2);
        \draw[thin, -latex] (dx2.east) -- (9.2,-2);
        \draw (9.2,0) circle (2pt);
        \draw (9.2,-1.5) circle (2pt);
        \draw[thin] (7.1,0) -- (9.1,0);
        \draw[thin] (7.1,-1.5) -- (9.1,-1.5);
        \draw[thin] (9.3,0) -- (9.7,0);
        \draw[thin] (9.3,-1.5) -- (9.7,-1.5);
        \node[outer sep=0pt,thin] at (10.1,0) {$\cdots$};
        \node[outer sep=0pt,thin] at (10.1,-1.5) {$\cdots$};
        \draw[thin] (10.5,0) -- (10.9,0);
        \draw[thin] (10.5,-1.5) -- (10.9,-1.5);
        \draw (11,0) circle (2pt);
        \draw (11,-1.5) circle (2pt);
        \node at (11,-0.75) (v13) {$v_{n-1}$};
        \draw[thin, -latex] (v13.north) -- (11,-0.1);
        \draw[thin, -latex] (v13.south) -- (11,-1.4);
        \draw (11.1,0) to [resistor] node [midway,above=1pt] {$r_{n,1}$} (12.3,0);
        \draw (12.3,0) to [inductor] node [midway,above=1pt] {$l_{n,1}$} (13.5,0);
        \draw (11.1,-1.5) to [resistor] node [midway,above=1pt] {$r_{n,2}$} (12.3,-1.5);
        \draw (12.3,-1.5) to [inductor] node [midway,above=1pt] {$l_{n,2}$} (13.5,-1.5);
        \draw[thin, -latex] (11.9,-0.3) -- node [midway,below=0.1] {$i_{n-1}$} (12.7,-0.3);
        \draw (13.5,0) to [resistor] node [midway,right=3pt] {$rb_{n}$} (13.5,-1.5);
        \draw (14.6,0) to [capacitor] node [midway,right=5pt] {$c_{n}$} (14.6,-1.5);
        \draw[thin] (11,-1.7) -- (11,-2.3);
        \draw[thin] (14.6,-1.7) -- (14.6,-2.3);
        \node at (12.8,-2) (dxn) {$\Delta x$};
        \draw[thin, -latex] (dxn.west) -- (11,-2);
        \draw[thin, -latex] (dxn.east) -- (14.6,-2);
        \draw[thin, -latex] (14.7,0.3) -- node [midway,above=0.1] {$i_{out}$} (15.4,0.3);
        \draw (15.5,0) to [resistor] node [midway,right=2pt] {$Z_{out}$} (15.5,-1.5);
        \node at (16.6,-0.75) (vout) {$v_{out}$};
        \draw[thin, -latex] (vout.north) -- (16.6,-0.1);
        \draw[thin, -latex] (vout.south) -- (16.6,-1.4);
        \draw (16.6,0) circle (2pt);
        \draw (16.6,-1.5) circle (2pt);
        \draw[thin] (13.5,0) -- (16.5,0);
        \draw[thin] (13.5,-1.5) -- (16.5,-1.5);
    \end{tikzpicture}
    \caption{Circuit network model with $n$ subsections to approximate a transmission line}
    \label{fig:fullModel}
\end{figure*}

The rest of this paper is organized as follows. \Cref{sec:tran} briefly reviews transmission line theory. \Cref{sec:cir} describes our method to approximate voltage and current along a transmission line through the circuit network model. To validate the correctness of our approximation result, we compare it to the transmission line theory in \Cref{sec:evenly} when the electrical attributes are distributed evenly. In \Cref{sec:unevenly}, we illustrate our method's capability of evaluating voltage and current along an unevenly distributed transmission line. That capability is illustrated by two examples. The first example is computing voltage and current along a railway track circuit when a ballast degradation occurs. The second one is assessing the current as a train is passing through an intact track circuit. Finally, \Cref{sec:con} concludes this paper.

\section{Transmission line theory}
\label{sec:tran}
In this section, we briefly review transmission line theory for a model shown in \Cref{fig:transModel}. The goal is to obtain the spatial and temporal distribution of voltage and current, \textit{i.e.} $v(x,t)$ and $i(x,t)$, given the boundary conditions at $x=0$ and the values of the electrical attributes listed in \Cref{tab:eleAtt}.
\begin{table}
    \centering
    \begin{tabular}{|c|c|c|}
        \hline
        $R$ & Series resistance & $\Omega/m$ \\\hline
        $L$ & Series inductance & $H/m$ \\\hline
        $G$ & Shunt conductance & $S/m$ \\\hline
        $C$ & Shunt capacitance & $F/m$ \\\hline
    \end{tabular}
    \caption{Notations of the electrical attributes used in the transmission line model}
    \label{tab:eleAtt}
\end{table}

For an evenly distributed transmission line, those electrical attributes are constant. Therefore, by using Kirchhoff's circuit laws, we have the following  within an infinitesimal distance $\Delta x$.
\begin{multline*}
    v(x+\Delta x,t)-R\Delta xi(x+\Delta x,t)
    -L\Delta x\frac{\partial i(x+\Delta x,t)}{\partial t}\\-v(x,t)=0,
    \end{multline*}
    and
    \[
    i(x+\Delta x,t)-G\Delta xv(x,t)-C\Delta x\frac{\partial v(x,t)}{\partial t}-i(x,t)=0.
\]
Taking $\Delta x\rightarrow0$, we obtain the following partial differential equations.
\begin{align*}
    &\frac{\partial v(x,t)}{\partial x}=Ri(x,t)+L\frac{\partial i(x,t)}{\partial t},\\
    &\frac{\partial i(x,t)}{\partial x}=Gv(x,t)+C\frac{\partial v(x,t)}{\partial t}.
\end{align*}
Using separation of variables, we assume
\begin{align*}
    v(x,t)&=Re\{v(x)e^{j(\omega t+\phi)}\}\\
    i(x,t)&=Re\{i(x)e^{j(\omega t+\phi)}\},
\end{align*}
which result in two decoupled second-order ordinary differential equations.
\begin{align*}
    \frac{d^2v}{dx^2}&=\gamma^2v,\\
    \frac{d^2i}{dx^2}&=\gamma^2i,
\end{align*}
where $\gamma=\sqrt{(R+j\omega L)(G+j\omega C)}$. Using the boundary conditions at $x=0$, the final results is
\begin{align}
    v(x,t)&=Re\{v(x)e^{j(\omega t+\phi)}\},\label{eq:tranRes1}\\
    i(x,t)&=Re\{i(x)e^{j(\omega t+\phi)}\},\label{eq:tranRes2}
\end{align}
where
\begin{align*}
    v(x)&=\frac{v(0)}{1+\mu}(e^{\gamma x}+\mu e^{-\gamma x}),\\
    i(x)&=\frac{i(0)}{1-\mu}(e^{\gamma x}-\mu e^{-\gamma x}),\\
    \mu&=\frac{Z_0-Z_c}{Z_0+Z_c},\\
    Z_c&=\sqrt{\frac{R+j\omega L}{G+j\omega C}}.
\end{align*}
Note that for an unevenly distributed transmission line, those electrical attributes may vary with the distance $x$, which makes such simple solutions difficult or impossible.

\section{Circuit network model}
\label{sec:cir}
In this section, we propose our procedure to quickly approximate the voltage and current along a transmission line which can have unevenly distributed physical parameters. Our method divides a long transmission line into $n$ subsections with equal length to form a circuit network where electrical attributes lump into each subsection as shown in \Cref{fig:fullModel}. The goal is to compute the voltage and current at every node connecting two adjacent subsections, \textit{i.e.} $v_g$ and $i_g$ in \Cref{fig:fullModel}. Those would be discrete approximations of the continuous results, $v(x,t)$ and $i(x,t)$ in \Cref{eq:tranRes1,eq:tranRes2}, given by transmission line theory.

It is clear that the electrical components in \Cref{fig:fullModel} can be either same or different, which incidentally does not affect the capability of our proposed method. When all components are the same, that is
\begin{align*}
    &r_{1,1}=r_{1,2}=r_{2,1}=r_{2,2}=\cdots=r,\\
    &l_{1,1}=l_{1,2}=l_{2,1}=l_{2,2}=\cdots=l,\\
    &rb_1=rb_2=\cdots=rb,\\
    &c_1=c_2=\cdots=c,
\end{align*}
the network is called undamaged, and those constants $r$, $l$, $rb$, and $c$ are the undamaged constants. Otherwise, the network is damaged, in which case we use a pair of two lists, $(\boldsymbol{l},\boldsymbol{e})$, to describe a specific damage case, where $\boldsymbol{l}$ is the list of damaged components, and $\boldsymbol{e}$ is the corresponding list of damage amounts. As a concrete example, the damage case
\begin{equation*}
    (\boldsymbol{l},\boldsymbol{e})=([rb_1,c_2],[0.1,2])
\end{equation*}
means $rb_1=0.1rb$, and $c_2=2c$, while all the other components are undamaged.

Our previous work \cite{Ni2020Frequency} proposed algorithms to compute frequency response and transfer functions for one-dimensional self-similar networks. For the specific application in this paper, we can use a recursive algorithm from \cite{Ni2020Frequency} to obtain the following two quantities at each node in \Cref{fig:fullModel},
\begin{enumerate}
    \item Impedance $Z_g=V_g/I_g$,
    \item Voltage gain $H_g=V_{out}/V_g$.
\end{enumerate}
That algorithm is listed in \Cref{alg:numFin}. The returned values \texttt{Z} and \texttt{H} are the impedance and the voltage gain at the transmitter end. The input argument \texttt{nG} is the number of subsections of the circuit network, and \texttt{w} is the angular frequency at which the computation conducts, which is same as the $\omega$ in \Cref{eq:tranRes1,eq:tranRes2}. In addition, \texttt{zOut} is the impedance at the receiver end. The undamaged constants $r$, $l$, $rb$, $c$, are grouped into the input argument \texttt{undCst}.
\begin{algorithm}
    \caption{Computing the impedance \texttt{Z} and voltage gain \texttt{H} at the angular frequency \texttt{w} for a circuit network with \texttt{nG} number of generations given its damage case \texttt{(l,e)} and the undamaged constants \texttt{undCst}}
    \label{alg:numFin}
    \begin{algorithmic}[1]
    \STATE{\texttt{\textbf{function}[Z,H]=fR(l,e,undCst,zOut,w,nG)}}
    \STATE{\texttt{s~=~j*w;}}
    \STATE{\texttt{[l1,e1,lS,eS]~=~partition(l,e);}}
    \STATE{\texttt{g1Cst~=~getG1Cst(l1,e1,undCst);}}
    \IF{\texttt{nG~==~0}}
    \STATE{\texttt{[Z,H]~=~G0(zOut,s);}}
    \ELSE
    \STATE{\texttt{nG~=~nG-1;}}
    \STATE{\texttt{[ZS,HS]=fR(lS,eS,undCst,zOut,w,nG);}}
    \STATE{\texttt{Z~=~Zr(g1Cst,ZS,s);}}
    \STATE{\texttt{H~=~Hr(g1Cst,Z,HS,s);}}
    \STATE{\texttt{save(Z,H);}}
    \ENDIF
    \end{algorithmic}
\end{algorithm}

\begin{figure*}
    \centering
    \begin{tikzpicture}[circuit ee IEC]
        \draw (0,0) circle (2pt);
        \draw (0,-1.5) circle (2pt);
        \node at (-0.4,0) {$+$};
        \node at (-0.4,-1.5) {$-$};
        \node at (0,-0.75) (vin) {$v_{in}$};
        \draw[thin, -latex] (vin.north) -- (0,-0.1);
        \draw[thin, -latex] (vin.south) -- (0,-1.4);
        \draw (0.1,0) to [resistor] node [midway,above=1pt] {$r_{1,1}$} (1.3,0);
        \draw (1.3,0) to [inductor] node [midway,above=1pt] {$l_{1,1}$} (2.5,0);
        \draw (0.1,-1.5) to [resistor] node [midway,above=1pt] {$r_{1,2}$} (1.3,-1.5);
        \draw (1.3,-1.5) to [inductor] node [midway,above=1pt] {$l_{1,2}$} (2.5,-1.5);
        \draw[thin, -latex] (0.8,-0.3) -- node [midway,below=0.1] {$i_{in}$} (1.7,-0.3);
        \draw (2.5,0) to [resistor] node [midway,right=3pt] {$rb_1$} (2.5,-1.5);
        \draw (3.7,0) to [capacitor] node [midway,right=5pt] {$c_1$} (3.7,-1.5);
        \draw[thin, -latex] (4.6,-0.3) -- node [midway,below=0.1] {$i_{1}$} (5.4,-0.3);
        \node at (5.7,-0.75) (v1) {$v_1$};
        \draw[thin, -latex] (v1.north) -- (5.7,-0.1);
        \draw[thin, -latex] (v1.south) -- (5.7,-1.4);
        \draw (5.7,0) circle (2pt);
        \draw (5.7,-1.5) circle (2pt);
        \draw[thin] (2.5,0) -- (5.6,0);
        \draw[thin] (2.5,-1.5) -- (5.6,-1.5);
        \node[draw,outer sep=0pt,thin] (Zs) at (6.5,-0.75) {$Z_s(s)$};
        \draw[thin] (5.8,0) -- (6.5,0);
        \draw[thin] (5.8,-1.5) -- (6.5,-1.5);
        \draw[thin] (Zs.north) -- (6.5,0);
        \draw[thin] (Zs.south) -- (6.5,-1.5);
    \end{tikzpicture}
    \caption{When coding the recursive \Cref{alg:numFin}, we only focus on the first generation, because the impedance of the subnetwork $Z_s(s)=V_1/I_1$ and the voltage gain of the subnetwork $H_s(s)=V_{out}/V_1$ are computed by the recursive call in \Cref{alg:numFin}. ($H_s(s)$ is not shown here.)}
    \label{fig:recModel}
\end{figure*}
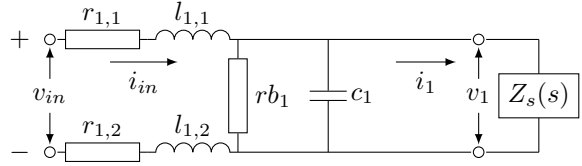

In \Cref{alg:numFin}, the \texttt{partition()} function splits the damage case (\texttt{l},\texttt{e}) for the entire network into two parts, where (\texttt{l1},\texttt{e1}) is the damage case at the first generation, and (\texttt{lS},\texttt{eS}) is the damage case with respect to the subnetwork after the first generation. Then, based on (\texttt{l1},\texttt{e1}), the \texttt{getG1Cst()} function computes the values of the constants at the first generation, that is the values of $r_{1,1}$, $r_{1,2}$, $l_{1,1}$, $l_{1,2}$, $rb_1$ and $c_1$. If the network in question has zero generations, the returned impedance $\texttt{Z}=\texttt{zOut}$, and the returned voltage gain $\texttt{H}=1$. Otherwise, the input argument \texttt{nG} is reduced by one, and a recursive call is made to obtain the impedance \texttt{ZS} and voltage gain \texttt{HS} for the subnetwork given the relevant damage case (\texttt{lS},\texttt{eS}). Those two quantities are used to compute the final result, the impedance \texttt{Z} and voltage gain \texttt{H} for the entire network. By using series and parallel connection rules of idealized electrical components, from \Cref{fig:recModel}, we can obtain that
\begin{align*}
    Z(s)&=r_{1,1}+r_{1,2}+l_{1,1}s+l_{1,2}s\\
    &+\cfrac{1}{\cfrac{1}{rb_1}+c_1s+\cfrac{1}{Z_s(s)}},\\
    H(s)&=H_s(s)(1-\cfrac{r_{1,1}+r_{1,2}+l_{1,1}s+l_{1,2}s}{Z(s)}),
\end{align*}
which are conducted by the \texttt{Zr()} and \texttt{Hr()} functions. Finally, the resultant \texttt{Z} and \texttt{H} are saved externally. Due to the recursive nature of \Cref{alg:numFin}, by doing so, all $Z_g$ and $H_g$ can be saved externally for any node $g$ between two neighboring subsections. Then, since $V_{out}$ is known, which serves as the boundary condition $v(0,t)$ in \Cref{sec:tran}, we can evaluate $V_g$ and $I_g$ at every node too.

Note that \Cref{alg:numFin} is not limited to the circuit network shown in \Cref{fig:fullModel}. For other similar networks, the structure of \Cref{alg:numFin} stays the same with necessary modifications on the detailed computations inside some subordinate functions. A more comprehensive explanation of \Cref{alg:numFin} can be found in our previous work \cite{Ni2020Frequency}.

\section{Evenly distributed electrical attributes}
\label{sec:evenly}
To prove the correctness of our calculation in \Cref{sec:cir}, we compare our discrete approximations to the continuous results given by the transmission line theory when electrical attributes are evenly distributed.

The boundary conditions at the receiver end are assumed to be
\begin{align}
    v(0,t)&=Re\{110e^{j4600\pi t}\}V,\label{eq:boundV}\\
    Z_0&=500\Omega,\nonumber\\
    i(0,t)&=Re\{0.22e^{j4600\pi t}\}A.\nonumber
\end{align}
The above voltage and frequency are taken from \cite{Wang2016Fault}. The length of the entire transmission line is set to be $1170m$, which is from \cite{Zhao2009The}. The constants for electrical attributes are from \cite{Hill1989Railway} at the frequency $2300Hz$, where
\begin{align*}
    R&=2.5m\Omega/m,\\
    L&=1.8\mu H/m,\\
    G&=20\mu S/m,\\
    C&=0.2nF/m.
\end{align*}
By knowing the above quantities, we can compute the voltage $v(x,t)$ and current $i(x,t)$ given by the transmission line theory according to \Cref{eq:tranRes1,eq:tranRes2}.

\begin{figure}
    \centering
    \includegraphics[width=.47\textwidth]{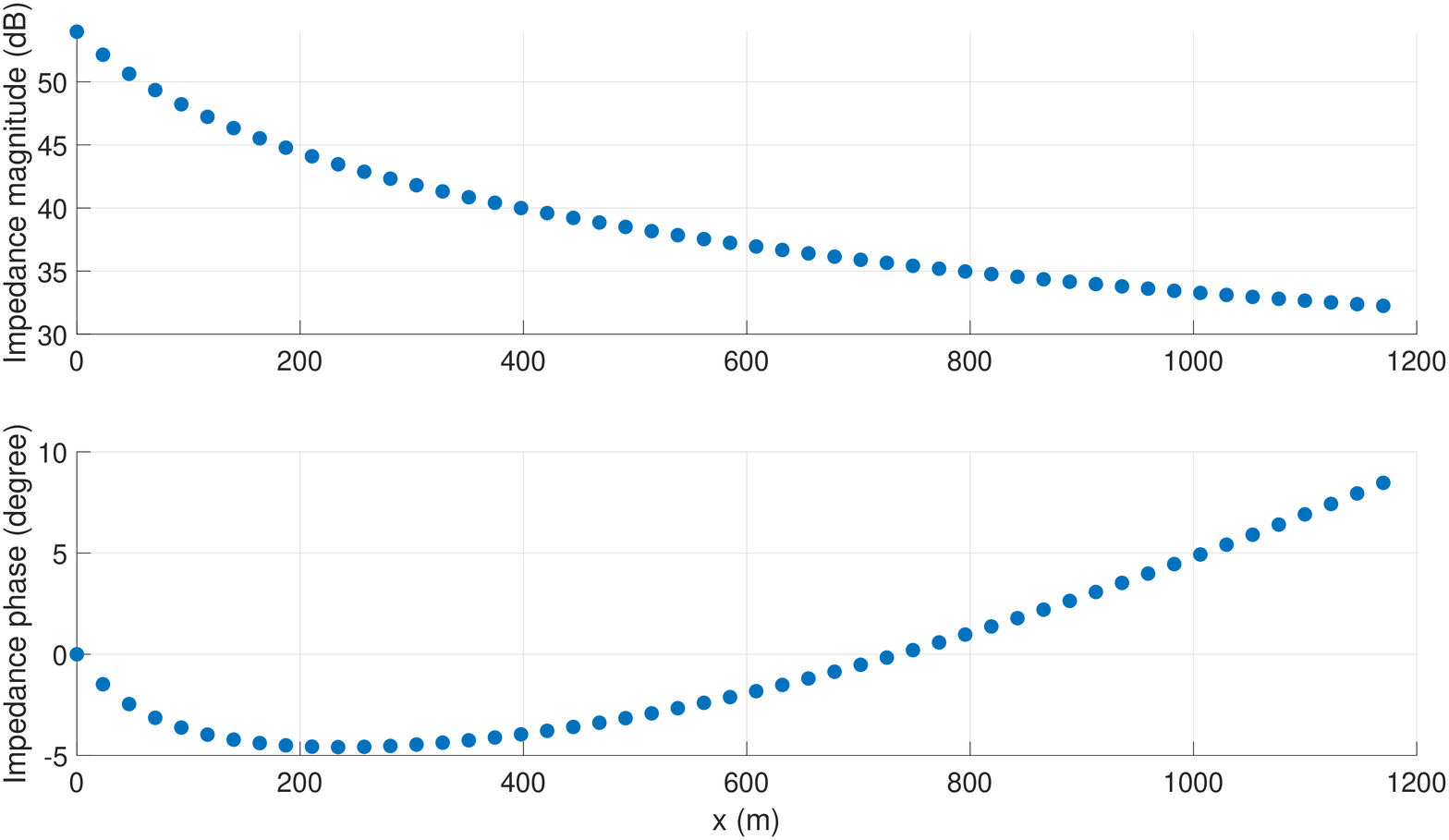}
    \caption{Impedance $Z$ at each node connecting two adjacent subsections obtained by the circuit network model}
    \label{fig:ZUndamaged}
\end{figure}

\begin{figure}
    \centering
    \includegraphics[width=.47\textwidth]{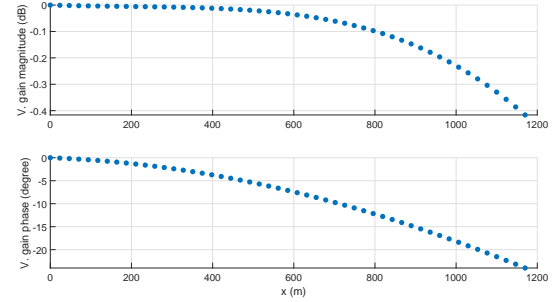}
    \caption{Voltage gain $H$ at each node connecting two adjacent subsections obtained by the circuit network model}
    \label{fig:HUndamaged}
\end{figure}

On the other hand, for our circuit network model, if we use $n$ subsections, the distance of each one is $\Delta x=1170/n$ meters. Therefore, the undamaged constants are
\begin{align*}
    r&=2.5\Delta x/2~m\Omega,\\
    l&=1.8\Delta x/2~\mu H,\\
    rb&=1/(20\times10^{-6}\Delta x)~\Omega,\\
    c&=0.2\Delta x~nF,
\end{align*}
which are grouped into the \texttt{undCst} to call \Cref{alg:numFin}. In addition, both \texttt{l} and \texttt{e} are empty lists indicating the intact case, $\texttt{zOut}=500$, $\texttt{w}=4600\pi$, and $\texttt{nG}=n$. Then, \Cref{alg:numFin} saves the impedance and voltage gain at each node between two adjacent subsections, which are shown in \Cref{fig:ZUndamaged,fig:HUndamaged} for the undamaged network with fifty generations. After that, given the same boundary condition $v(0,t)$ in \Cref{eq:boundV}, we can obtain the voltage and current at those nodes.

\begin{figure}
    \centering
    \includegraphics[width=.47\textwidth]{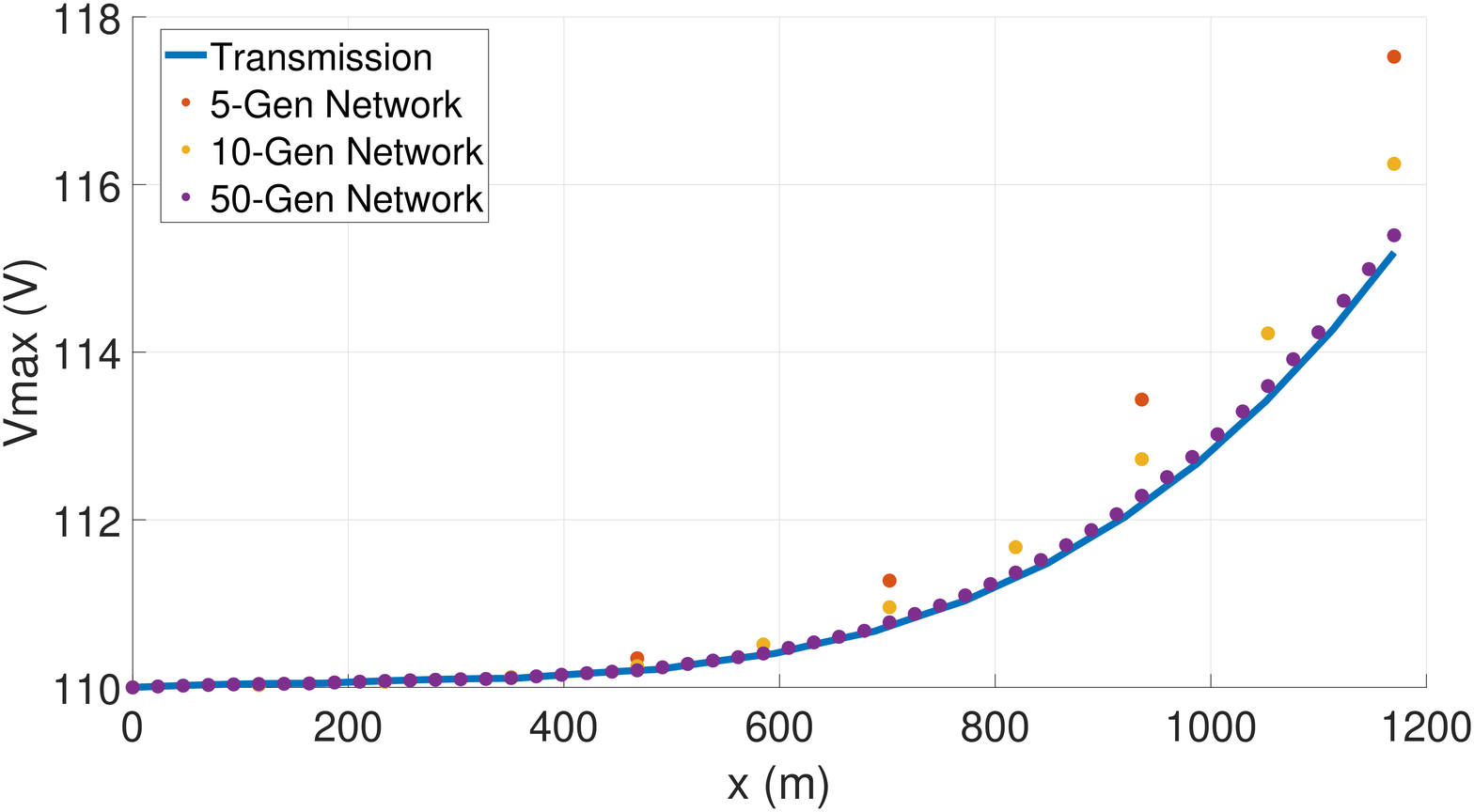}
    \caption{Maximum voltage $\max_t(|v(x,t)|)$ \textit{versus} the distance $x$. The blue curve is given by transmission line theory. The dots are given by our circuit network model}
    \label{fig:vMaxUndamaged}
\end{figure}

\begin{figure}
    \centering
    \includegraphics[width=.47\textwidth]{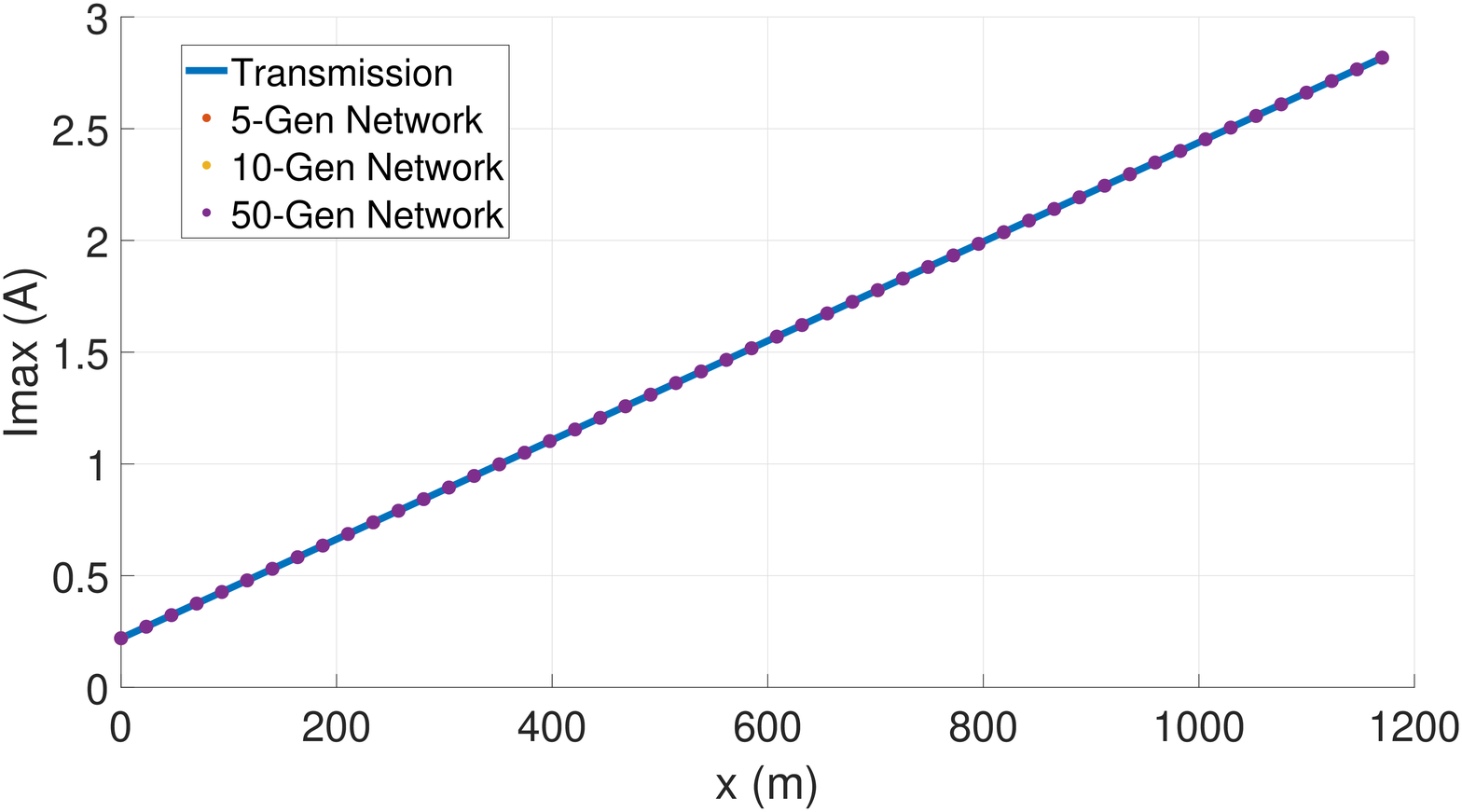}
    \caption{Maximum current $\max_t(|i(x,t)|)$ \textit{versus} the distance $x$. The blue curve is given by transmission line theory. The dots are given by our circuit network model}
    \label{fig:iMaxUndamaged}
\end{figure}

\begin{figure}
    \centering
    \includegraphics[width=.47\textwidth]{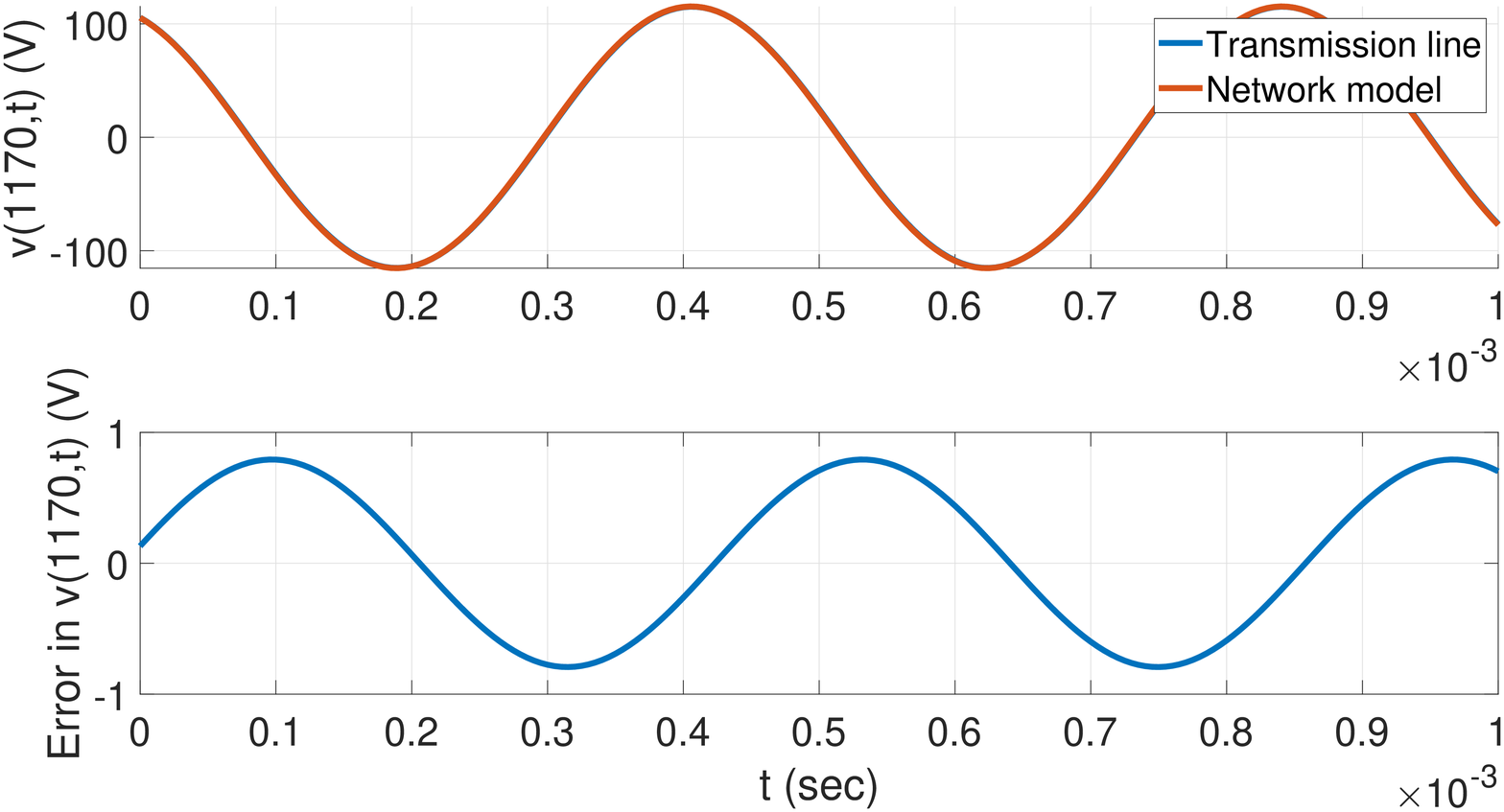}
    \caption{The upper figure plots $v(1170,t)$ obtained by two methods. The blue curve is from transmission line theory, while the red curve is from our network model with fifty generations, where two curves almost overlap each other. The lower figure shows the error between those two}
    \label{fig:vUndamaged}
\end{figure}

The comparison between the transmission line theory's result and our circuit network's result are shown in \Cref{fig:vMaxUndamaged,fig:iMaxUndamaged}, where $\max_t(|v(x,t)|)$ and $\max_t(|i(x,t)|)$ are plotted \textit{versus} the distance $x$. For the circuit network, we test three networks with $5$, $10$, and $50$ generations. From \Cref{fig:vMaxUndamaged}, we see that the $V_{\max}$ given by the circuit model converges to the one obtained by using transmission line theory as the number of generations increases. From \Cref{fig:iMaxUndamaged}, we observe that the $I_{\max}$ given by both methods almost overlap each other. In addition, to show that both magnitudes and phases are correct, \Cref{fig:vUndamaged} compares $v(1170,t)$ at the receiver end given by both methods. Hence, we can confirm that our method provides a reasonable approximation of transmission line model.

\section{Unevenly distributed electrical attributes}
\label{sec:unevenly}
In this section, we illustrate our method's capability of computing voltage and current along an unevenly distributed transmission line with the applications to railway track circuit networks. All constants used in this section are same as those in \Cref{sec:evenly}. Besides, we fix the number of subsections at $117$ here, so each subsection takes $10m$. In addition, we assume $V_{max}$ at the transmitter end is fixed at the one in \Cref{fig:vMaxUndamaged}, \textit{i.e.} $V_{max}=115V$ at $x=1170m$.

\begin{figure}
    \centering
    \includegraphics[width=.47\textwidth]{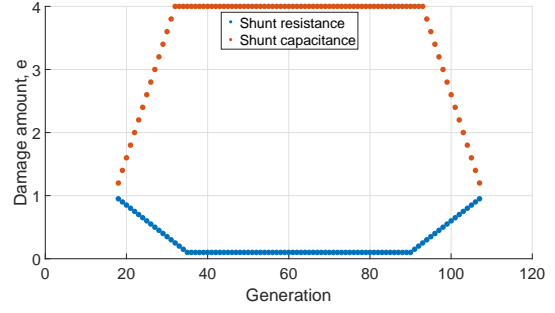}
    \caption{The damage amounts $e$ are plotted \textit{versus} generations where the blue dots are for the shunt resistance from $rb_{18}$ to $rb_{107}$, and the red dots are for the shunt capacitance from $c_{18}$ to $c_{107}$}
    \label{fig:eDamaged1}
\end{figure}

\begin{figure}
    \centering
    \includegraphics[width=.47\textwidth]{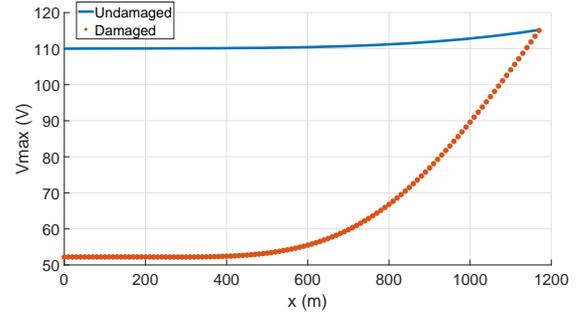}
    \caption{The distribution of $\max_t(|v(x,t)|)$ at each node between two adjacent subsections obtained by the circuit network model when the ballast degradation occurs. The undamaged curve is obtained by transmission line theory which is same as that in \Cref{fig:vMaxUndamaged}}
    \label{fig:vMaxDamaged1}
\end{figure}

\begin{figure}
    \centering
    \includegraphics[width=.47\textwidth]{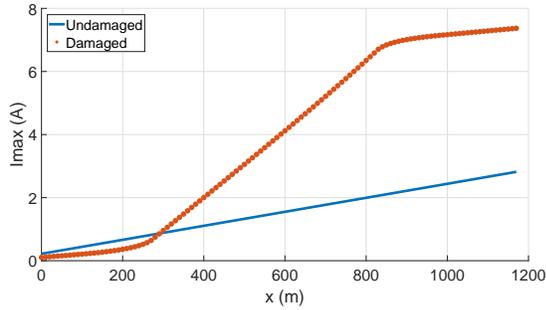}
    \caption{The distribution of $\max_t(|i(x,t)|)$ at each node between two adjacent subsections obtained by the circuit network model when the ballast degradation occurs. The undamaged curve is obtained by transmission line theory which is same as that in \Cref{fig:iMaxUndamaged}}
    \label{fig:iMaxDamaged1}
\end{figure}

The first example of unevenly distribution we showcase here is ballast degradation, which means unusual current leakage between the rails through the ballast \cite{Bruin2017Railway}. In this damage case, some shunt resistance $rb_g$ become lower and shunt capacitance $c_g$ become higher. Here, we suppose a ballast degradation happens between $x=100m$ and $x=1000m$, that is between subsection $18$ and subsection $107$. Note that the order of subsections is opposite to the direction of $x$. Hence, the list of damaged components is
\begin{equation*}
    \boldsymbol{l}=\begin{bmatrix}rb_{18}&\cdots&rb_{107}&c_{18}&\cdots&c_{107}\end{bmatrix}.
\end{equation*}
The assumed list of damage amounts $e$ is plotted in \Cref{fig:eDamaged1}.
When the above damage case $(\boldsymbol{l},\boldsymbol{e})$ is inputted to \Cref{alg:numFin}, the resultant voltage and current distribution along the track obtained by our circuit network model for this type of ballast degradation are shown in \Cref{fig:vMaxDamaged1,fig:iMaxDamaged1}. Note that the other two damage cases mentioned in \cite{Bruin2017Railway}, insulation imperfections and rail conductance impairments, can also be simulated by our circuit network model.

\begin{figure}
    \centering
    \includegraphics[width=.47\textwidth]{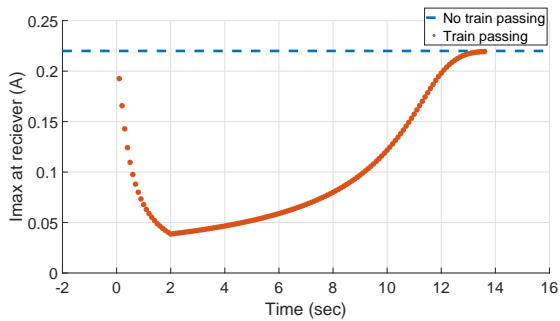}
    \caption{The variation in the current at the receiver end, $\max_t|i(0,t)|$, as the train passing through an intact track obtained by our modeling algorithm}
    \label{fig:iMaxDamaged2}
\end{figure}

The second example of uneven distribution is when a train is moving along an intact track, which can be equivalently viewed as a number of damage cases. We assume that a $190$-meter train is moving from the receiver end to the transmitter end at a constant speed $100m/s$, that is it passes one subsection every $0.1sec$. Besides, the train has one wheel base every $10$ meters, so there are $20$ wheel bases in total. Each wheel base acts as an additional shunt resistor across two rails with resistance $r_w=102.0408\Omega$. The undamaged value of the shunt resistance in this case is
\begin{equation*}
    rb=1/(20\times10^{-6}\times10)=5000\Omega.
\end{equation*}
When the shunt resistance $rb$ is connected to the wheel base's resistance $r_w$ in parallel, the equivalent resistance is $100\Omega$. In other words, when one wheel base is within one subsection, we can consider that as if that corresponding shunt resistance is damaged by a factor of $0.02$. Therefore, this example of a train moving along an intact track can be regarded as a time series of damage cases, where the correspondence between the time instance and the damage case is listed in \Cref{tab:train}.
\begin{table}
    \centering
    \begin{tabular}{|c|c|}
        \hline
        Time (sec) & Damage case $(\boldsymbol{l},\boldsymbol{e})$ \\
        \hline
        0.1 & $([rb_{117}],[0.02])$ \\
        0.2 & $([rb_{117},rb_{116}],[0.02,0.02])$ \\
        $\vdots$ & $\vdots$ \\
        2 & $([rb_{117},\cdots,rb_{98}],[0.02,\cdots,0.02])$ \\
        2.1 & $([rb_{116},\cdots,rb_{97}],[0.02,\cdots,0.02])$ \\
        $\vdots$ & $\vdots$ \\
        11.7 & $([rb_{20},\cdots,rb_{1}],[0.02,\cdots,0.02])$ \\
        11.8 & $([rb_{19},\cdots,rb_{1}],[0.02,\cdots,0.02])$ \\
        $\vdots$ & $\vdots$ \\
        13.6 & $([rb_1],[0.02])$ \\
        \hline
    \end{tabular}
    \caption{The equivalent damage case at each time instance for the train passing example}
    \label{tab:train}
\end{table}
At each time instance in \Cref{tab:train}, we use the corresponding damage case in the second column to call \Cref{alg:numFin}. Then, we can evaluate the current at the receiver end, $\max_t|i(0,t)|$ as the train is passing through this sector of track, which is plotted in \Cref{fig:iMaxDamaged2}. Note that \Cref{fig:iMaxDamaged2} shares similar characteristics with real measurements presented in \cite{Bruin2017Railway}.

It is worth pointing out that every damage case only takes around one second to compute with the setup of the $117$-generation network in this section. The computation is run on \texttt{MATLAB R2019b} with a single CPU of Intel Core i7-4510U Processor.

\section{CONCLUSIONS}
\label{sec:con}
This paper presents a method to rapidly and accurately compute voltage and current along a transmission line, especially when its electrical properties are unevenly distributed. The proposed method divides a transmission line into many subsections to form a circuit network, where electrical attributes are lumped at each generation. Thus, that model offers opportunities to imitate an unevenly distributed transmission line by a damaged network. Thanks to a modular recursive algorithm designed to compute frequency response for a general class of one-dimensional self-similar networks, we can thereby quantify the voltage and current at each node inside that network. The correctness of our method is proved by comparing our result to the one given by the transmission line theory when the electrical attributes are evenly distributed. In addition, we illustrate our method's ability to simulate an unevenly distributed transmission line by providing two real application examples of railway track circuit systems. One of them concerns a damage situation when a ballast degradation occurs. The other quantifies the current's variation as a train is passing through. The proposed method of quickly simulating a railway track circuit under various conditions can be applicable to the relevant health monitoring research area. For instance, it may be suitable for some damage detection methods with supervised learning tools and offers a baseline case of how the circuit would respond with respect to different types of situations.

\bibliographystyle{IEEEtran}
\bibliography{references}

\begin{thebibliography}{10}
\providecommand{\url}[1]{#1}
\csname url@rmstyle\endcsname
\providecommand{\newblock}{\relax}
\providecommand{\bibinfo}[2]{#2}
\providecommand\BIBentrySTDinterwordspacing{\spaceskip=0pt\relax}
\providecommand\BIBentryALTinterwordstretchfactor{4}
\providecommand\BIBentryALTinterwordspacing{\spaceskip=\fontdimen2\font plus
\BIBentryALTinterwordstretchfactor\fontdimen3\font minus
  \fontdimen4\font\relax}
\providecommand\BIBforeignlanguage[2]{{%
\expandafter\ifx\csname l@#1\endcsname\relax
\typeout{** WARNING: IEEEtran.bst: No hyphenation pattern has been}%
\typeout{** loaded for the language `#1'. Using the pattern for}%
\typeout{** the default language instead.}%
\else
\language=\csname l@#1\endcsname
\fi
#2}}

\bibitem{Wilson2012Grounding}
\BIBentryALTinterwordspacing
P.~Wilson, ``Chapter 1 - grounding and wiring,'' in \emph{The Circuit
  Designer's Companion (Third Edition)}, P.~Wilson, Ed.\hskip 1em plus 0.5em
  minus 0.4em\relax Oxford: Newnes, 2012, pp. 1 -- 43. [Online]. Available:
  \url{http://www.sciencedirect.com/science/article/pii/B978008097138400001X}
\BIBentrySTDinterwordspacing

\bibitem{Rouphael2014Antenna}
\BIBentryALTinterwordspacing
T.~J. Rouphael, ``Chapter 1 - antenna systems, transmission lines, and matching
  networks,'' in \emph{Wireless Receiver Architectures and Design}, T.~J.
  Rouphael, Ed.\hskip 1em plus 0.5em minus 0.4em\relax Boston: Academic Press,
  2014, pp. 1 -- 60. [Online]. Available:
  \url{http://www.sciencedirect.com/science/article/pii/B9780123786401000019}
\BIBentrySTDinterwordspacing

\bibitem{Lampe2016Power}
\BIBentryALTinterwordspacing
L.~Lampe and L.~Berger, ``Chapter 16 - power line communications,'' in
  \emph{Academic Press Library in Mobile and Wireless Communications}, S.~K.
  Wilson, S.~Wilson, and E.~Biglieri, Eds.\hskip 1em plus 0.5em minus
  0.4em\relax Oxford: Academic Press, 2016, pp. 621 -- 659. [Online].
  Available:
  \url{http://www.sciencedirect.com/science/article/pii/B9780123982810000168}
\BIBentrySTDinterwordspacing

\bibitem{Poljak2019Simplified}
\BIBentryALTinterwordspacing
D.~Poljak and M.~Cvetković, ``Chapter 4 - simplified models of the human
  body,'' in \emph{Human Interaction with Electromagnetic Fields}, D.~Poljak
  and M.~Cvetković, Eds.\hskip 1em plus 0.5em minus 0.4em\relax Academic
  Press, 2019, pp. 91 -- 122. [Online]. Available:
  \url{http://www.sciencedirect.com/science/article/pii/B9780128164433000121}
\BIBentrySTDinterwordspacing

\bibitem{Hill1993Rail}
R.~J. {Hill} and D.~C. {Carpenter}, ``Rail track distributed transmission line
  impedance and admittance: theoretical modeling and experimental results,''
  \emph{IEEE Transactions on Vehicular Technology}, vol.~42, no.~2, pp.
  225--241, 1993.

\bibitem{Wang2016Fault}
Z.~{Wang}, J.~{Guo}, Y.~{Zhang}, and R.~{Luo}, ``Fault diagnosis for jointless
  track circuit based on intrinsic mode function energy moment and optimized
  ls-svm,'' in \emph{2016 IEEE International Conference on High Voltage
  Engineering and Application (ICHVE)}, 2016, pp. 1--4.

\bibitem{Zhao2009The}
L.~{Zhao}, J.~{Guo}, H.~{Li}, and W.~{Liu}, ``The simulation analysis of
  influence on jointless track circuit signal transmission from compensation
  capacitor based on transmission-line theory,'' in \emph{2009 3rd IEEE
  International Symposium on Microwave, Antenna, Propagation and EMC
  Technologies for Wireless Communications}, 2009, pp. 1113--1118.

\bibitem{Verbert2015Exploiting}
K.~{Verbert}, B.~{De Schutter}, and R.~{Babuška}, ``Exploiting spatial and
  temporal dependencies to enhance fault diagnosis: Application to railway
  track circuits,'' in \emph{2015 European Control Conference (ECC)}, 2015, pp.
  3047--3052.

\bibitem{Hill1989Railway}
R.~J. {Hill}, D.~C. {Carpenter}, and T.~{Tasar}, ``Railway track admittance,
  earth-leakage effects and track circuit operation,'' in \emph{Proceedings.,
  Technical Papers Presented at the IEEE/ASME Joint Railroad Conference}, 1989,
  pp. 55--62.

\bibitem{Chen2008Fault}
\BIBentryALTinterwordspacing
J.~Chen, C.~Roberts, and P.~Weston, ``Fault detection and diagnosis for railway
  track circuits using neuro-fuzzy systems,'' \emph{Control Engineering
  Practice}, vol.~16, no.~5, pp. 585 -- 596, 2008. [Online]. Available:
  \url{http://www.sciencedirect.com/science/article/pii/S0967066107001244}
\BIBentrySTDinterwordspacing

\bibitem{Wang2010Modeling}
S.~guo {Wang} and B.~{Wang}, ``Modeling of distributed rlc interconnect and
  transmission line via closed forms and recursive algorithms,'' \emph{IEEE
  Transactions on Very Large Scale Integration (VLSI) Systems}, vol.~18, no.~1,
  pp. 119--130, 2010.

\bibitem{Ni2020Frequency}
X.~Ni and B.~Goodwine, ``Frequency response and transfer functions of large
  self-similar networks,'' 2020.

\bibitem{Bruin2017Railway}
T.~{de Bruin}, K.~{Verbert}, and R.~{Babuška}, ``Railway track circuit fault
  diagnosis using recurrent neural networks,'' \emph{IEEE Transactions on
  Neural Networks and Learning Systems}, vol.~28, no.~3, pp. 523--533, 2017.

\end{thebibliography}

\addtolength{\textheight}{-12cm}   % This command serves to balance the column lengths

%%%%%%%%%%%%%%%%%%%%%%%%%%%%%%%%%%%%%%%%%%%%%%%%%%%%%%%%%%%%%%%%%%%%%%%%%%%%%%%%

\end{document}